\documentclass[journal,twoside,web]{ieeecolor}
\usepackage{generic}
\usepackage{cite}
\usepackage{amsmath,amssymb,amsfonts}
\usepackage{algorithmic}
\usepackage{algorithm}
\usepackage{graphicx}
\usepackage{amsmath}
\usepackage{booktabs}
\usepackage{graphicx}
\usepackage{times}
\usepackage{soul}
\usepackage{multirow}
\usepackage{booktabs}       
\usepackage{booktabs}       
\usepackage{amsfonts}       
\usepackage{nicefrac}       
\usepackage{microtype}      
\usepackage{amsmath}
\usepackage{graphicx}
\usepackage{colortbl}
\usepackage{pifont}

\definecolor{gray}{rgb}{0.5,0.5,0.5}
\definecolor{gray94}{gray}{.92}
\definecolor{gray90}{gray}{.90}
\definecolor{gray85}{gray}{.85}

\newcommand{\gray}[1]{\textcolor{gray}{#1}}
\definecolor{gray}{rgb}{0.5,0.5,0.5}
\definecolor{darkergreen}{RGB}{21, 152, 56}
\definecolor{gray95}{gray}{.95}
\definecolor{gray90}{gray}{.90}
\usepackage{etoolbox}
\usepackage{listings}
\makeatletter
\AfterEndEnvironment{algorithm}{\let\@algcomment\relax}
\AtEndEnvironment{algorithm}{\kern2pt\hrule\relax\vskip3pt\@algcomment}
\let\@algcomment\relax
\newcommand\algcomment[1]{\def\@algcomment{\footnotesize#1}}
\renewcommand\fs@ruled{\def\@fs@cfont{\bfseries}\let\@fs@capt\floatc@ruled
  \def\@fs@pre{\hrule height.8pt depth0pt \kern2pt}%
  \def\@fs@post{}%
  \def\@fs@mid{\kern2pt\hrule\kern2pt}%
  \let\@fs@iftopcapt\iftrue}
\makeatother

\begin{document}

\title{Neuro-BERT: Rethinking Masked Autoencoding for Self-supervised Neurological Pretraining}

\author{Di Wu, Siyuan Li, Jie Yang, \IEEEmembership{Member, IEEE}, and Mohamad Sawan, \IEEEmembership{Fellow, IEEE}
\thanks{
Di Wu, Jie Yang, and Mohamad Sawan are with the Center of Excellence in Biomedical Research on Advanced Integrated-on-chips Neurotechnologies (CenBRAIN Neurotech), School of Engineering, Westlake University, Hangzhou 310024, China. (e-mail: yangjie@westlake.edu.cn; sawan@westlake.edu.cn).} 
\thanks{ Siyuan Li is with the School of Engineering, Westlake University, Hangzhou 310024, China.}
\thanks{
 Mohamad Sawan is also with the Westlake Institute for Optoelectronics, Fuyang, Hangzhou 311421, China.}
\thanks{This work was supported by STI2030-Major Projects (2022ZD0208805), National Natural Science Foundation of China (Grant No. 623B2085), "Pioneer" and "Leading Goose" R\&D Program of Zhejiang (2024C03002), and the Key Project of Westlake Institute for Optoelectronics (Grant No. 2023GD004).}
}

\maketitle

\begin{abstract}

Deep learning associated with neurological signals is poised to drive major advancements in diverse fields such as medical diagnostics, neurorehabilitation, and brain-computer interfaces. The challenge in harnessing the full potential of these signals lies in the dependency on extensive, high-quality annotated data, which is often scarce and expensive to acquire, requiring specialized infrastructure and domain expertise. To address the appetite for data in deep learning, we present \textbf{Neuro-BERT}, a self-supervised pre-training framework of neurological signals based on masked autoencoding in the Fourier domain. The intuition behind our approach is simple: frequency and phase distribution of neurological signals can reveal intricate neurological activities. We propose a novel pre-training task dubbed \textbf{Fourier Inversion Prediction (FIP)}, which randomly masks out a portion of the input signal and then predicts the missing information using the Fourier inversion theorem. Pre-trained models can be potentially used for various downstream tasks such as sleep stage classification and gesture recognition. Unlike contrastive-based methods, which strongly rely on carefully hand-crafted augmentations and siamese structure, our approach works reasonably well with a simple transformer encoder with no augmentation requirements. By evaluating our method on several benchmark datasets, we show that \textbf{Neuro-BERT} improves downstream neurological-related tasks by a large margin.

\end{abstract}

\section{Introduction}
The ubiquity and versatility of neurological signals have recently rendered them an essential area of study with far-reaching implications in various research fields~\cite{cao2018data,otter2020survey}. Of these signals, surface electromyography (sEMG) demonstrates promising potential in interpreting user intentions from human muscle activities. \textcolor{black}{This capability facilitates the development of novel command interfaces in numerous human-machine interaction (HMI) scenarios, enhancing user interaction experiences~\cite{li2018dynamic, teng2020design}, enabling more natural and immersive interactions~\cite{reidy2020facial}, and rehabilitation~\cite{mulas2005emg}.}  \textcolor{black}{Additionally, other neurological signals, such as electroencephalography (EEG), are pivotal in medical applications, playing a vital role in areas like disease diagnosis~\cite{wu2022bridging, wu2021c}.}

\begin{figure}[t]
\centering
  \includegraphics[width=1.0\linewidth, bb=0 0 560 360]{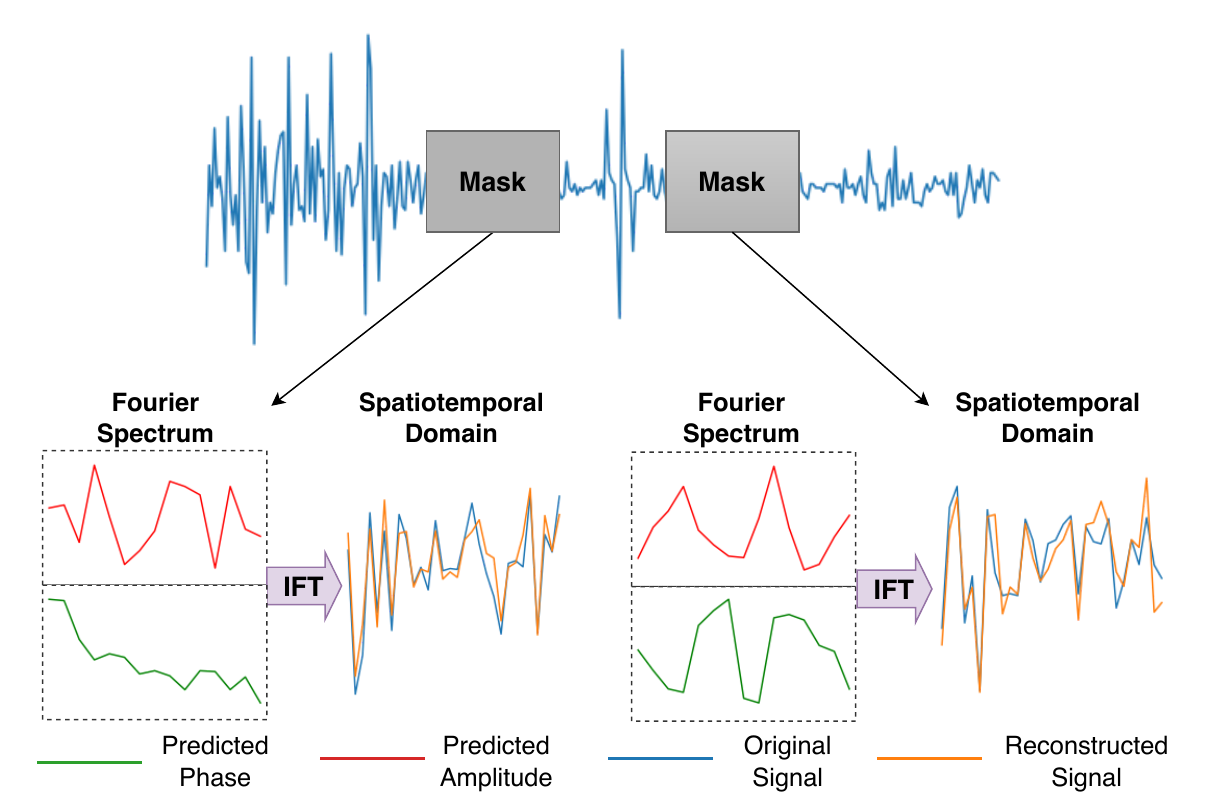}
  \caption{Example of the proposed Fourier inversion prediction on EMG signal. The model predicts the phase and amplitude of the Fourier spectrum and reconstructs the missing signal segments using the inverse Fourier transform (IFT). For each masked signal segment, we illustrate the predicted phase and amplitude components of the Fourier spectrum as well as the corresponding original and reconstructed signal.}
  \label{fig_intro}
\end{figure}

\textcolor{black}{However, the advancements in utilizing these signals have predominantly relied on supervised deep learning methods, which require extensive labeled datasets not always available in real-world settings. Consequently, learning representations without supervision through pretext tasks has become increasingly popular. In fields like computer vision, early self-supervised learning methods aimed to capture invariant features by predicting image transformations~\cite{eccv2016coloring,iclr2018rotation}. These methods, however, do not readily translate to other modalities due to their reliance on vision-specific heuristics. More recently, self-supervised contrastive learning approaches witnessed significant progress, even outperforming supervised methods on several downstream vision tasks.~\cite{chen2020simple, nips2020byol, caron2020unsupervised}. Similarly, in natural language processing, techniques such as autoregressive language modeling~\cite{radford2018improving} and masked autoencoding~\cite{devlin2018bert} have driven significant advancements by focusing on predicting masked tokens~\cite{yang2021multi, wang-etal-2018-glue, 10.1145/1390156.1390177}.}

However, self-supervised pre-training for neurological signals is far from established. \textcolor{black}{Unlike human languages and natural images, neurological signals are notably informatively sparse, which complicates the task of learning generic representations through self-supervised training.} Inspired by successes in other domains, some researchers have adapted contrastive-based pre-training methods to signals like EEG. These methods typically depend on assumptions that do not translate robustly outside of visual contexts, such as the requirement for strong, semantically consistent augmentations~\cite{ijcai2021-324, 9414752, wu2021align}.\textcolor{black}{Inspired by the success in other domains, some researchers have proposed contrastive-based pre-training methods for signals like EEG, relying on assumptions that may not hold as strongly outside of visual contexts, such as the need for strong, semantically consistent augmentations~\cite{ijcai2021-324, 9414752, wu2021align}. For instance, the \textit{RandomResizedCrop} (RRC) is fundamental in vision contrastive learning, but no equivalent cropping and resizing-based augmentation exists for neurological signals. The most commonly adopted augmentation technique for neurological signals is bandpass filtering, where a particular band of frequencies is removed from the original signal. Unfortunately, the filtered signal is not guaranteed to be semantically consistent with the original signal, as a useful frequency band might be filtered out.} Other augmentations, such as adding random noise and channel flipping, are weak augmentations that bring limited improvements over the learned representations. Moreover, augmentations for neurological signals require deliberate hyper-parameter tuning due to the variation of different neurological signals. As reported in \cite{ijcai2021-324, 9414752}, the hyper-parameters of augmentations vary greatly across tasks and datasets.

\textcolor{black}{To address these challenges, we propose Neuro-BERT,  a novel self-supervised framework tailored for neurological signals. Drawing inspiration from the success of BERT~\cite{devlin2018bert}, Neuro-BERT utilizes a unique approach by predicting the frequency and phase of masked neurological signal segments and reconstructing the original signal using the Fourier inversion theorem. This method allows the model to gain a deeper understanding of the underlying neurological activities. For example, the frequency of an EEG signal is a critical indicator of brain activity, while the phase of an EMG signal reflects patterns of muscle fiber recruitment. An illustration of the proposed self-supervised neurological pre-training is given in Fig. \ref{fig_intro}. Our approach is intuitive and practically straightforward. Compared to contrastive-based methods that strongly rely on carefully designed and tuned data augmentations, Neuro-BERT works reasonably well without any augmentations, thereby avoiding any potential corruption of the original data. Additionally, Neuro-BERT does not require the siamese network architecture commonly used in many contrastive-based methods~\cite{chen2020simple, nips2020byol, caron2020unsupervised}. This simplifies the model structure while maintaining robust performance, making it particularly suitable for practical applications in neurological research.}

Our contributions are summarized as follows:
\begin{itemize}
    \item We modify existing self-supervised pre-training approaches proposed for CV and NLP to accommodate neurological signals and establish them as strong baselines to alleviate reliance on massive data annotation.
    \item We further present Neuro-BERT, a novel self-supervised pre-training framework specifically designed for neurological signals.
    \item We propose a tailored pre-training task, FIP, which predicts the missing information via the Fourier inversion theorem. We demonstrate that models pre-trained with FIP better capture the underlying neurological activities than naive imputation in the spatiotemporal domain.
    \item Comprehensive experiments on both EEG and EMG benchmarks show that our proposed pre-training framework achieves new state-of-the-art performances.
\end{itemize}

\section{Related Work}
\subsection{Contrastive Learning}
Contrastive learning methods learn instance-level discriminative representations by extracting invariant features over distorted views of the same data point. MoCo~\cite{cvpr2020moco} adopted a large memory bank to introduce enough informative negative samples, while SimCLR~\cite{chen2020simple} adopted a larger batch size to replace the memory bank mechanism. BYOL variants~\cite{nips2020byol, chen2020simsiam,chen2021empirical,iccv2021dino} and Barlow Twins variants~\cite{zbontar2021barlow} further eliminate the requirement of negative samples, using various techniques to avoid representation collapse. Motivated by the success of contrastive learning in CV, some research endeavors are made to adopt contrastive learning onto time series~\cite{9414752, ijcai2021-324}. TS-TCC~\cite{ijcai2021-324} and CoST~\cite{iclr2022CoST} proposed to contrast time series from both temporal and contextual domains using different augmentation techniques on different views of the signal. In particular, they randomly add noise and re-scale one view of the signal while segmenting, rearranging, and re-scaling the other view as data augmentation. Similarly, SSL-EEG~\cite{banville2021uncovering} involves three pre-text tasks: relative positioning, temporal shuffling, and contrastive predictive coding. Due to the high variability of neurological signals, applying contrastive learning directly to neurological representation learning inevitably introduces additional hyper-parameters making the augmentation-related parameter tuning a troublesome task. The natural clustering property induced by pulling near similar representations and pushing away dissimilar samples makes the learned representations linearly separable. Linear probing~\cite{chen2020simple} is therefore commonly adopted to evaluate the performances of contrastive-based methods. However, it misses the opportunity to learn strong but non-linear features, which limits the power of deep learning. Thus, in this work, we aim to maximize the performance of the model on downstream tasks with the fine-tuning protocol~\cite{chen2020simsiam, bao2021beit, he2021masked}.

\subsection{Masked Autoencoding}
The pioneering work of applying the notion of autoregressive modeling to learn representations is the classic autoencoding~\cite{hinton1993autoencoders}. Autoencoders first map the input data to a latent space and then reconstruct the input from the representation in the latent space. Denoising autoencoders~\cite{vincent2010stacked} are a family of autoencoders that reconstruct the uncorrupted input signal with a corrupted version of the signal as input. Generalizing the denoising autoregressive modeling, masked predictions attracted the attention of both the NLP and vision community. BERT performs masked language modeling, the task of which is to predict the randomly masked input tokens. Representations learned by BERT as pre-training generalize well to various downstream tasks.
For CV, inpainting tasks~\cite{pathak2016context} to predict large missing regions using convolutional networks is proposed to learn representation. iGPT~\cite{chen2020generative} predicts succeeding pixels given a sequence of pixels as input. \textcolor{black}{MaskedAE}~\cite{he2021masked} and BEiT~\cite{bao2021beit} mask out random patches of the input image and reconstruct the missing patches with a Vision Transformer (ViT). Only visible patches are fed to the encoder in \textcolor{black}{MaskedAE}, while both visible and masked patches are fed in BEiT. The most appealing property of autoregressive modeling, compared to contrastive-based methods, is its simplicity and minimal reliance on dataset-specific augmentation engineering.

\begin{figure*}[t]
    \centering 
    \includegraphics[width=0.91\linewidth]{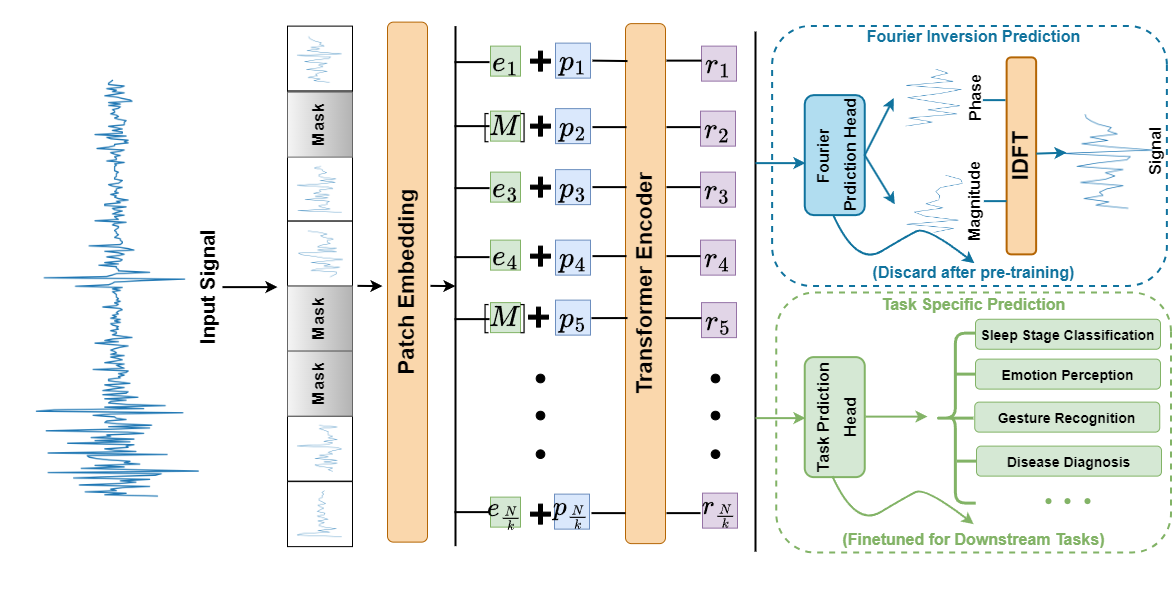}
    \vspace{-1.5em}
    \caption{Illustration of Neuro-BERT for neurological self-supervised pre-training. \textcolor{black}{The input neurological signal is segmented into non-overlapping frames and transformed into a sequence of patch embeddings $e$ using a 1-D convolutional layer, where $N$ denotes the signal length and $k$ the convolution kernel size. Each patch embedding is added with a learnable positional embedding $p$ before being processed by the transformer encoder. The encoder output $r$ serves as the neurological patch representation. During self-supervised pre-training, a subset of these patches is randomly masked—represented by $[M]$ for the learnable mask token—and the model is trained to predict their frequency and phase distribution. The original signal is then reconstructed using the inverse discrete Fourier transform (IDFT).} The Fourier prediction head is discarded after pre-training, and a linear prediction head will be adopted for various classification or regression downstream tasks on top of the pre-trained model.}
    \label{sytem_fig}
\end{figure*}
\section{Method}
\subsection{Masked Autoencoding Preliminaries}
\label{sec:pre}
Our framework performs the masked autoencoding task as pre-training, which reconstructs the masked neurological content up to some details. A typical masked autoencoding framework contains the following five main components:
\begin{enumerate}
    \item Patch embedding (tokenizer): In analogous to the common practice in NLP where each word in a sentence is first mapped to a word embedding vector, an input neurological signal is mapped to a sequence of tokens with each token containing information within a time window.
    \item Masking strategy: Given the neurological token sequence, different masking strategies design the rule of which tokens to mask and how to mask the tokens. After masking, the transformed token sequence is passed to the encoder. 
    \item Encoder architecture: The encoder extracts latent representations from the masked neurological token sequence to recover the masked part of the original signal or certain features of the original signal. Besides recovering, the learned representations are also expected to be generic for various downstream tasks.
    \item Prediction target: The prediction target defines what form of the original signal to predict. It can either be the magnitude of the signal in the spatiotemporal domain or some transformation of the signal.
    \item Prediction head architecture (decoder): The prediction head takes the latent representation from the encoder and produces certain forms of the original signal at the masked locations.
\end{enumerate}
\subsection{Neuro-BERT}
We will elaborate on the specific designs and innovations of our proposed Neuro-BERT in relation to the previously mentioned components. A general illustration of the proposed Neuro-BERT is shown in Fig.~\ref{sytem_fig}.
\paragraph{\textcolor{black}{Patch Embedding}}\textcolor{black}{Instead of directly linearly projecting the input data to the model dimension, we use non-overlapping 1-D convolution to encode the input signal into a sequence of segments following the work of Vision Transformers (ViT)~\cite{dosovitskiy2020image}. Formally, let $x \in \mathbb{R}^{N \times C}$ be a neurological signal of length $N$ with $C$ channels, a non-overlapping 1-D convolution with kernel size $k$ is applied on $x$ to produce a sequence of $\frac{N}{k}$ $d$-dimensional patch embeddings $[e_1,e_2,...,e_{\frac{N}{k}}]$. The resulting embeddings could also be viewed as a vector map $E \in \mathbb{R}^{\frac{N}{k} \times d}$ where $d$ is the projection dimension. The convolution operation is followed by a GELU non-linearity~\cite{hendrycks2016bridging}.}
\label{patch_emd}
\textcolor{black}{\paragraph{Masking Strategy}Given the sequence of patch embeddings $[e_1,e_2,...,e_{\frac{N}{k}}]$ of a neurological signal embedded by the 1-D convolution, we randomly mask a subset of the embeddings with a masking ratio $r$ following a uniform distribution. During pre-training, the masked tokens are replaced and initialized with a learnable token of the same dimension $d$ as the visible tokens. Since the transformer is a feed-forward architecture that does not capture the input temporal order information, we add a learnable positional embedding to each patch embedding. The processed sequence of embeddings is then fed into the transformer encoder.}
\label{mask}
\paragraph{Encoder Architecture}\textcolor{black}{Similar to the field of vision, convolution-based architectures~\cite{acmmm2021eeg, lawhern2018eegnet,ijcai2021-324} are dominant in the existing domain of neurological applications.} \textcolor{black}{These architectures are often tailored with specific network structures and data flow to address modality-specific tasks. However, the inherent spatiotemporal inductive bias of convolution operations restricts CNN-based architectures from seamlessly integrating advanced techniques like token masking \cite{devlin2018bert} and positional embedding \cite{vaswani2017attention}, which are pivotal in masked autoencoding. To overcome these limitations, we utilize a transformer-based architecture that provides a unified end-to-end framework capable of processing raw neurological signals. Our architecture follows the standard transformer configuration with minor modifications to suit our specific application better.} The architecture includes a patch embedding layer followed by four blocks of multi-head self-attention (MSA). A detailed description of our encoder architecture is provided in Sec.~\ref{exp:exp}.
\label{encoder_arch}
\paragraph{Prediction Targets}
\label{prediction_task}
The no-sweat approach would be to directly reconstruct the amplitude of the masked neurological signal in the spatiotemporal domain following the standard practice~\cite{bao2021beit, he2021masked}. However, unlike images and natural languages, which contain sophisticated semantic information, inpainting masked neurological signals in the spatiotemporal domain is challenging due to their stochastic, nonstationary, and nonlinear nature~\cite{moss2004stochastic}. To tackle this issue, we propose \textbf{Fourier Inversion Prediction (FIP)} as a means of establishing a connection between neurological signals and the underlying neural activities. Take EEG and EMG as an example. EEG signal is usually divided into bandwidths known as delta, theta, alpha, and beta, which are associated with neurological activities~\cite{kirmizi2006comparative, jackson2014neurophysiological}. For instance, the alpha band (8-13 Hz) is associated with brain maturity and inhibition control, while the beta band (13-30 Hz) is linked to motor behavior. On the other hand, motor units, the basic contraction elements in the human motor system, could be viewed as filters that are often categorized by amplitude and phase response. While the muscle is contracted, it moves longitudinally and transversely based on the fiber direction~\cite{heckman2004physiology}. The bio-impedance of the motor unit as a filter thus varies, leading to a varying amplitude and phase response.

With these bio-inspired motivations in mind, we formalize the proposed FIP as follows. Firstly, given a discrete neurological signal $x$ of channel $c$ and length $N$, the discrete Fourier transform (DFT) is defined by:
\begin{equation}
\begin{aligned}
\textcolor{black}{X_{m}} = \sum_{n=1}^{n=N} 
x_n * e^{-\frac{2\pi j}{N}mn},
\end{aligned}
\label{eq1}
\end{equation}
where $m \in [1, N]$. We further expand Equation \eqref{eq1} into the real and imaginary parts using Euler's formula:
\begin{equation}
\begin{aligned}
\textcolor{black}{X_{m}} = \sum_{n=1}^{n=N} 
 \underbrace{x_n * \cos(\frac{2\pi}{N}mn)}_{\mathrm{real}}-\underbrace{j * \sin(\frac{2\pi}{N}mn)}_{\mathrm{imaginary}},
\end{aligned}
\label{eq2}
\end{equation}
where $j$ is the imaginary unit satisfying $j^2 = -1$. Specifically, \textcolor{black}{$X_{m}$} represents the spectrum of the sequence $x_n$ at the frequency $\omega_m = 2\pi m/N$. We can then calculate the magnitude $\left\|\textcolor{black}{X_{m}}\right\|^2$ and phase $\theta_{m}$:
\begin{equation}
\begin{aligned}
\left\|\textcolor{black}{X_{m}}\right\|^2 = \frac{1}{N} \sqrt{\mathrm{Re}(X_m)^2 + \mathrm{Im}(X_m)^2}\\
\theta_{m} = \mathrm{atan2}(\mathrm{Re}(X_m)^2, \mathrm{Im}(X_m)^2),
\end{aligned}
\label{eq3}
\end{equation}
\begin{algorithm}[t]
\caption{Pseudocode of Fourier Inversion Prediction.}
\label{alg:code}
\algcomment{\fontsize{7.2pt}{0em}\selectfont \texttt{rfft/irfft}: 1D FFT/IFFT for real signal
}
\definecolor{codeblue}{rgb}{0.25,0.5,0.5}
\lstset{
  backgroundcolor=\color{white},
  basicstyle=\fontsize{7.2pt}{7.2pt}\ttfamily\selectfont,
  columns=fullflexible,
  breaklines=true,
  captionpos=b,
  commentstyle=\fontsize{7.2pt}{7.2pt}\color{codeblue},
  keywordstyle=\fontsize{7.2pt}{7.2pt},
}
\begin{lstlisting}[language=python]
# x: masked patch sequence, B x N x C 
# X: the discrete Fourier transform of B x (N/2+1) x C
# model : the transformer model 
X = rfft2(x, dim=1)
magnitude, phase = model(x)
X_tilde = magnitude * torch.exp(1j * phase)
denoised_x = irfft2(X_tilde, dim=1)
\end{lstlisting}
\end{algorithm}
where $\mathrm{atan2}(\cdot, \cdot)$ is the arctangent function, $\mathrm{Re}(\cdot)$ and $\mathrm{Im}(\cdot)$ indicate real and imaginary parts respectively. During the pre-training process, the model is first pre-trained to predict the missing amplitude and phase of the signal. \textcolor{black}{Then the inverse DFT (IDFT) is applied to transform the predicted amplitude $\left\|\textcolor{black}{\hat{X}_{m}}\right\|^2$ and phase $\textcolor{black}{\hat{\theta_{m}}}$ back into the spatiotemporal domain:
\begin{equation}
\begin{aligned}
\textcolor{black}{\hat{X}_{m}} &= \left\|\textcolor{black}{\hat{X}_{m}}\right\|^2 * e^{j * \textcolor{black}{\hat{\theta_{m}}}} \\
\textcolor{black}{\hat{x}_{n}} &= \sum_{m=1}^{m=N} \textcolor{black}{\hat{X}_{m}} * e^{-\frac{2\pi j}{N}mn}.
\end{aligned}
\label{eq4}
\end{equation}}

\textcolor{black}{The loss adopted during pre-training minimizes the mean square error (MSE) loss between the inversely transformed signal  $\hat{x}$ and the original version of the masked signal $x$.} It is worth noting that for real neurological signal input $x$, its DFT is conjugate symmetric, which implies that half of the DFT contains complete information about the frequency and phase characteristics of $x$. This reduces the prediction target size by half compared to directly predicting missing information from the spatiotemporal domain. Finally, DFT computation is cheap and introduces negligible overhead by the use of fast Fourier transform (FFT) algorithms, which take advantage of the symmetry and periodicity properties of the Fourier transform. \textcolor{black}{We provide a pseudo code in PyTorch style in Algorithm~\ref{alg:code}.}

\paragraph{Prediction Head Architecture}\textcolor{black}{During the pre-training phase, the prediction head (decoder) is solely utilized to estimate the magnitude and phase of the underlying signal at masked locations. Therefore, the decoder architecture can be of arbitrary form independent of the encoder architecture, provided it effectively performs the masked autoencoding task. Our proposed Neuro-BERT adopts a linear layer as the prediction head, which is extremely computationally lightweight. The structure of this linear layer decoder is detailed in Section \ref{exp:exp}. Furthermore, we demonstrate in Sec. \ref{ablation:decoder} that this simplified linear layer decoder achieves commendable performance relative to more complex decoder architectures.}
\label{decoder_arch}

\section{Experiments}
\label{exp:exp}
We briefly introduce the datasets used in this work in Section \ref{exp:datasets} and provide training details in Section \ref{exp:implementation}. Then, we provide the results of the main experiment in Section \ref{exp:results}, and lastly, we provide ablation studies in Section~\ref{exp:ablation}.

\subsection{Benchmarks}
\label{exp:datasets}
\paragraph{Epileptic Seizure Detection} An epileptic seizure is a period of symptoms resulting from abnormally excessive or synchronized neuronal activity in the brain, which can be detected using EEG signals. The Epileptic Seizure Recognition Dataset~\cite{andrzejak2001indications} contains EEG recordings from 500 subjects consisting of healthy subjects and subjects suffering from epilepsy. For Each subject, 23.6 seconds of brain activity EEG is recorded, then segmented into one-second chunks with annotated labels. The original dataset provides five categories of annotations: EEG recorded with eyes open, EEG recorded with eyes closed, EEG recorded from healthy brain regions, EEG recorded from tumor brain regions, and EEG recorded during seizure onset. We follow most works where the previous four categories are merged and classified against the last category of seizure onset.

\paragraph{Sleep Stage Recognition} Human sleep can be divided into the following five stages: Wake (W), Non-rapid eye movement (N1, N2, N3), and Rapid Eye Movement (REM). The Sleep-EDF Dataset~\cite{physiobank2000physionet} contains EEG recordings from Caucasian males and females (21–35 years old) without any medication. The EEG signal is required at a sampling rate of 100 HZ. Following previous studies~\cite{eldele2021attention, ijcai2021-324}, we utilize the Fpz-Cz channel in this study. The EEG recordings are segmented into 10-second windows.

\paragraph{Hand Gesture Recognition} Humans perform finger and wrist movements by contracting muscles of the forearm. The Ninapro DB5 Dataset~\cite{pizzolato2017comparison} contains EMG signals collected from 10 intact subjects using 16 active single–differential wireless electrodes. Each subject wears two armbands close to the elbow during data acquisition. DB5 contains three exercises for each subject: (1) Basic finger movements, (2) Isometric and isotonic hand movements, as well as basic wrist movements, and (3) Grasping and functional movements. During each exercise, the subjects are required to repeat each movement six times with rest in between to avoid muscle fatigue. In our experiment, we randomly pick five repetitions as the training set and the remaining one as the testing set. We choose the exercise (2), which contains 17 different gestures within this work. The EMG recordings are segmented using a sliding window method with window size 50 and stride five.

\paragraph{Finger Joint Angle Regression} For the task of finger joint angle regression, we also use the Ninapro DB5 Dataset~\cite{pizzolato2017comparison}. Besides categorical labels of all three exercises in DB5, each subject is asked to wear a data glove to collect finger joint kinematic information during EMG acquisition. There are, in total, 22 sensors placed on the glove. Since the abduction and abduction movement of the fingers is associated with muscle located in the palm, we removed these sensors from this study and only chose sensors that reflect the finger joint angle during the extension and flexion movement of the fingers. In particular, we pick the sensors placed at The distal interphalangeal (DIP) joint and metacarpophalangeal (MCP) for the thumb and sensors placed at DIP, proximal interphalangeal (PIP), and MCP for the remaining four fingers. The joint angle data is normalized for each sensor, respectively. We adopt exercise (1) for this task, which contains 12 fine finger gestures. The recordings are segmented the same way we described for the hand gesture recognition task. We calculate and adopt the mean of the joint angle for each EMG window as prediction labels.

\begin{table}[t]
    \caption{Detailed architecture specifications for ConvNet and Transformer backbones utilized in this work.}
    \centering
\addtolength{\tabcolsep}{-6pt}
\begin{tabular}{c|c|c}
\toprule
& ConvNet  & Transformer \\ \hline
Stem & K$\times$1, 32, stride 4 & P$\times$1, 128, stride P \\ \hline
\multirow{3}{*}{Block 1} & 
\multirow{3}{*}{$\begin{bmatrix}\text{7$\times$1, 64}\\\text{MaxPool, stride 2}\end{bmatrix}$ $\times$ 1}  & 
\multirow{3}{*}{$\begin{matrix}\begin{bmatrix}\text{MSA, 128, rel. pos.}\\\text{\quad1$\times$1, 512\quad}\\\text{\quad1$\times$1, 128\quad}\end{bmatrix}\end{matrix}$  $\times$ 1} \\
& & \\
& & \\
\hline
\multirow{3}{*}{Block 2} & 
\multirow{3}{*}{$\begin{bmatrix}\text{7$\times$1, 128}\\\text{MaxPool, stride 2}\end{bmatrix}$ $\times$ 1}  & 
\multirow{3}{*}{$\begin{matrix}\begin{bmatrix}\text{MSA, 128}\\\text{\quad1$\times$1, 512\quad}\\\text{\quad1$\times$1, 128\quad}\end{bmatrix}\end{matrix}$  $\times$ 1} \\
& & \\
& & \\
\hline
\multirow{3}{*}{Block 3} & 
\multirow{3}{*}{$\begin{bmatrix}\text{7$\times$1, 256}\\\text{MaxPool, stride 2}\end{bmatrix}$ $\times$ 1}  & 
\multirow{3}{*}{$\begin{matrix}\begin{bmatrix}\text{MSA, 128}\\\text{\quad1$\times$1, 512\quad}\\\text{\quad1$\times$1, 128\quad}\end{bmatrix}\end{matrix}$  $\times$ 1} \\
& & \\
& & \\
\hline
\multirow{3}{*}{Block 4} & 
\multirow{3}{*}{$\begin{bmatrix}\text{7$\times$1, 512}\end{bmatrix}$ $\times$ 1}  & 
\multirow{3}{*}{$\begin{matrix}\begin{bmatrix}\text{MSA, 128}\\\text{\quad1$\times$1, 512\quad}\\\text{\quad1$\times$1, 128\quad}\end{bmatrix}\end{matrix}$  $\times$ 1} \\
& & \\
& & \\
\hline
\# params.
&
$1.22 \times 10^6$ (K=4)
&
$0.798 \times 10^6$ (P=4) \\
    \bottomrule
    \end{tabular}
    \label{tab:exp_network}
    \vspace{-1.0em}
\end{table}

\begin{table*}[t]
    \setlength{\tabcolsep}{0.9mm}
    \caption{Comparison with existing self-supervised methods on SleepEDF, Epilepsy, and Ninapro datasets. MSE, MAE, Acc (\%), and macro F1 (\%) for fine-tuning evaluations are reported.}
    \vspace{-0.5em}
    \centering
\resizebox{1.0\linewidth}{!}{
\begin{tabular}{l|cc|cc|cc|cc|cc}
\toprule
                                         &                &             & \multicolumn{2}{c|}{SleepEDF}                                       & \multicolumn{2}{c|}{Epilepsy}                                       & \multicolumn{2}{c|}{Ninapro (Classification)}                       & \multicolumn{2}{c}{Ninapro (Regression)}                           \\
Methods                                  & Pre-training   & Backbone    & FT (Acc)$\uparrow$               & FT (F1)$\uparrow$                & FT (Acc)$\uparrow$               & FT (F1)$\uparrow$                & FT (Acc)$\uparrow$               & FT (F1)$\uparrow$                & FT(MSE)$\downarrow$             & FT(MAE)$\downarrow$              \\ \hline
Rand. init.                              & Random         & ConvNet     & 83.34$\pm$0.42                   & 76.71$\pm$0.65                   & 97.46$\pm$0.26                   & 95.61$\pm$0.42                   & 91.21 $\pm$0.52                  & 80.35 $\pm$0.89                  & 5.61 $\pm$0.54                  & 63.38 $\pm$2.57                  \\
Sup. baseline                            & Supervised     & ConvNet     & 84.23$\pm$0.56                   & 77.84$\pm$0.74                   & 97.35$\pm$0.31                   & 94.54$\pm$0.57                   & 89.15 $\pm$0.67                  & 78.96 $\pm$0.78                  & 5.73 $\pm$0.66                  & 63.32 $\pm$2.63                  \\
SimCLR~\cite{chen2020simple}             & Contrastive    & ConvNet     & 83.60$\pm$0.79                   & 76.96$\pm$0.89                   & 97.52$\pm$0.41                   & 95.98$\pm$0.76                   & 91.73 $\pm$0.75                  & 80.86 $\pm$0.67                  & 5.54 $\pm$0.57                  & 63.50 $\pm$1.64                  \\
BYOL~\cite{nips2020byol}                 & Contrastive    & ConvNet     & 83.29$\pm$0.76                   & 76.68$\pm$0.94                   & 97.94$\pm$0.64                   & 96.70$\pm$0.87                   & 91.56 $\pm$0.84                  & 80.69 $\pm$0.95                  & 5.72 $\pm$0.45                  & 63.74 $\pm$1.97                  \\
SwAV~\cite{caron2020unsupervised}        & Contrastive    & ConvNet     & 83.84$\pm$1.35                   & 77.18$\pm$1.62                   & 98.60$\pm$1.12                   & 97.82$\pm$0.53                   & 91.84 $\pm$1.76                  & 81.07 $\pm$2.45                  & 5.78 $\pm$0.65                  & 63.32 $\pm$2.20                  \\
MoCo.V3~\cite{chen2021empirical}         & Contrastive    & ConvNet     & 83.46$\pm$0.68                   & 76.70$\pm$0.76                   & 98.74$\pm$0.58                   & 97.98$\pm$0.64                   & 92.14 $\pm$0.66                  & 81.45 $\pm$0.78                  & 5.46 $\pm$0.39                  & 62.83 $\pm$1.92                  \\
TS-TCC~\cite{ijcai2021-324}              & Contrastive    & ConvNet     & 84.05$\pm$0.82                   & 77.04$\pm$0.88                   & 98.75$\pm$0.78                   & 97.81$\pm$0.72                   & 92.02 $\pm$0.95                  & 80.93 $\pm$1.12                  & 5.14 $\pm$0.40                  & 62.95 $\pm$1.83                  \\
CoST~\cite{iclr2022CoST}                 & Contrastive    & ConvNet     & 84.21$\pm$0.45                   & 77.95$\pm$0.68                   & 98.87$\pm$0.83                   & 97.93$\pm$0.76                   & 92.19 $\pm$0.81                  & 81.96 $\pm$0.93                  & 5.56 $\pm$0.43                  & 63.47 $\pm$1.96                  \\ \hline
Rand. init.                              & Random         & Transformer & 84.20$\pm$0.41                   & 77.15$\pm$0.59                   & 97.88$\pm$0.27                   & 96.54$\pm$0.38                   & 91.97 $\pm$0.58                  & 81.46 $\pm$0.56                  & 5.28 $\pm$0.48                  & 61.21 $\pm$1.93                  \\
MoCo.V3~\cite{chen2021empirical}         & Contrastive    & Transformer & 85.49$\pm$0.72                   & 77.73$\pm$0.80                   & 98.93$\pm$0.59                   & 98.60$\pm$0.58                   & 92.15 $\pm$0.71                  & 83.23 $\pm$0.72                  & 5.57 $\pm$0.51                  & 62.38 $\pm$2.04                  \\
\rowcolor{gray95}\textcolor{black}{MaskedAE}~\cite{he2021masked} & Autoregressive & Transformer & 85.08$\pm$0.89                   & 77.40$\pm$0.86                   & 99.15$\pm$0.74                   & 98.79$\pm$0.64                   & 93.03 $\pm$0.78                  & 84.32 $\pm$0.82                  & 5.03 $\pm$0.45                  & 60.79 $\pm$1.71                  \\
\rowcolor{gray95}\textcolor{black}{GPT~\cite{chen2020generative}} & \textcolor{black}{Autoregressive} & \textcolor{black}{Transformer} & \textcolor{black}{84.56$\pm$0.77}                   & \textcolor{black}{77.26$\pm$0.92}                   & \textcolor{black}{98.22$\pm$0.49}                   & \textcolor{black}{97.12$\pm$0.78}                   & \textcolor{black}{92.26 $\pm$0.84}                  & \textcolor{black}{83.13 $\pm$0.56}                  & \textcolor{black}{5.15 $\pm$0.23}                  & \textcolor{black}{61.20 $\pm$1.57}                  \\
\rowcolor{gray95}\bf{Neuro-BERT}          & Autoregressive & Transformer & \textbf{86.53}$\pm$\textbf{0.83} & \textbf{78.94}$\pm$\textbf{0.87} & \textbf{99.34}$\pm$\textbf{0.66} & \textbf{98.82}$\pm$\textbf{0.65} & \textbf{94.28}$\pm$\textbf{0.72} & \textbf{86.69}$\pm$\textbf{0.78} & \textbf{4.72}$\pm$\textbf{0.43} & \textbf{60.11}$\pm$\textbf{1.46} \\
\bottomrule
\end{tabular}
    }
    \label{tab:exp_comp_ft}
    \vspace{-0.5em}
\end{table*}

\begin{table*}[t]
    \setlength{\tabcolsep}{1.0mm}
    \caption{Comparison with existing self-supervised methods on SleepEDF, Epilepsy, and Ninapro datasets. Acc (\%) and mAP (\%) for linear evaluations are reported.}
    \vspace{-0.5em}
    \centering
\resizebox{1.0\linewidth}{!}{
\begin{tabular}{l|cc|cc|cc|cc}
\toprule
                                        &                &             & \multicolumn{2}{c|}{SleepEDF}             & \multicolumn{2}{c|}{Epilepsy}             & \multicolumn{2}{c}{Ninapro (Classification)} \\
Methods                                  & Pre-training   & Backbone    & kNN (Acc)$\uparrow$ & SVM (mAP)$\uparrow$ & kNN (Acc)$\uparrow$ & SVM (mAP)$\uparrow$ & kNN (Acc)$\uparrow$   & SVM (mAP)$\uparrow$  \\ \hline
Rand. init.                              & Random         & ConvNet     & 65.82$\pm$4.96      & 69.15$\pm$4.14      & 90.13$\pm$3.56      & 97.52$\pm$4.35      & 64.69 $\pm$5.79       & 10.61 $\pm$3.43      \\
SimCLR~\cite{chen2020simple}             & Contrastive    & ConvNet     & 72.47$\pm$0.84      & 75.47$\pm$0.68      & 95.28$\pm$0.56      & 96.09$\pm$0.64      & 87.14 $\pm$0.82       & 65.86 $\pm$0.87      \\
BYOL~\cite{nips2020byol}                 & Contrastive    & ConvNet     & 71.65$\pm$0.73      & 75.26$\pm$0.81      & 94.54$\pm$0.48      & 96.64$\pm$0.52      & 85.57 $\pm$0.87       & 66.02 $\pm$0.96      \\
SwAV~\cite{caron2020unsupervised}        & Contrastive    & ConvNet     & \bf{76.19$\pm$1.24} & 76.83$\pm$1.56      & 95.95$\pm$1.23      & 98.36$\pm$1.43      & 86.13 $\pm$2.33       & 64.43 $\pm$2.12      \\
MoCo.V3~\cite{chen2021empirical}         & Contrastive    & ConvNet     & 72.53$\pm$0.75      & 75.51$\pm$0.84      & 94.33$\pm$0.31      & 95.09$\pm$0.34      & 87.45 $\pm$0.66       & 66.29 $\pm$0.78      \\
TS-TCC~\cite{ijcai2021-324}              & Contrastive    & ConvNet     & 73.87$\pm$0.76      & 76.12$\pm$0.98      & 96.98$\pm$0.35      & 98.74$\pm$0.37      & 86.87 $\pm$0.95       & 66.10 $\pm$0.93      \\ \hline
Rand. init.                              & Random         & Transformer & 66.75$\pm$4.10      & 69.50$\pm$3.26      & 90.23$\pm$3.74      & 96.29$\pm$4.03      & 65.28 $\pm$4.73       & 11.07 $\pm$4.51      \\
MoCo.V3~\cite{chen2021empirical}         & Contrastive    & Transformer & 75.61$\pm$0.63      & \bf{77.84$\pm$0.82} & \bf{98.17$\pm$0.34} & \bf{98.91$\pm$0.42} & \bf{89.01 $\pm$0.71}  & \bf{69.34 $\pm$0.74} \\
\rowcolor{gray95}\textcolor{black}{MaskedAE}~\cite{he2021masked} & Autoregressive & Transformer & 68.43$\pm$0.89      & 70.78$\pm$0.94      & 90.64$\pm$0.49      & 96.95$\pm$0.59      & 77.53 $\pm$0.84       & 48.67 $\pm$0.82      \\
\rowcolor{gray95}\textcolor{black}{GPT~\cite{chen2020generative}} & \textcolor{black}{Autoregressive} & \textcolor{black}{Transformer} & \textcolor{black}{67.29$\pm$0.75}     & \textcolor{black}{70.05$\pm$0.72}      & \textcolor{black}{90.86$\pm$0.43}      & \textcolor{black}{96.33$\pm$0.64}      & \textcolor{black}{74.96 $\pm$0.67}       & \textcolor{black}{46.27 $\pm$0.84}      \\
\rowcolor{gray95}\bf{Neuro-BERT}          & Autoregressive & Transformer & 69.55$\pm$0.82      & 71.49$\pm$0.84      & 93.78$\pm$0.47      & 97.06$\pm$0.57      & 79.38 $\pm$0.82       & 53.25 $\pm$0.84      \\
\bottomrule
\end{tabular}}
    \label{tab:exp_comp_linear}
    \vspace{-0.5em}
\end{table*}

\begin{table}[th]
\centering
    \setlength{\tabcolsep}{0.5mm}
    \caption{Description of used datasets in our experiments.}
    \vspace{-4pt}
\resizebox{\columnwidth}{!}{
    \begin{tabular}{l|cccc}
    \toprule
Datasets           & SleepEDF                                           & Epilepsy                                            & \multicolumn{2}{c}{Ninapro}                        \\
                   & \multicolumn{1}{c}{\cite{physiobank2000physionet}} & \multicolumn{1}{c}{\cite{andrzejak2001indications}} & \multicolumn{2}{c}{\cite{pizzolato2017comparison}} \\ \hline
Downstream Task    & Classification                                     & Classification                                      & Classification             & Regression            \\
Train samples      & 25612                                              & 9200                                                & 118403                     & 118403                \\
Test samples       & 8910                                               & 2300                                                & 39468                      & 39468                 \\
Class number       & 5                                                  & 2                                                   & 18                         & -                     \\
Sequence length    & 3000                                               & 178                                                 & 50                         & 50                    \\
Channel number     & 1                                                  & 1                                                   & 16                         & 16                    \\ \hline
ConvNet            & K=25                                               & K=4                                                 & \multicolumn{2}{c}{K=8}                            \\
Transformer        & P=30                                               & P=4                                                 & \multicolumn{2}{c}{P=4}                            \\ \hline
Random Scaling     & $\sigma=1.5$                                       & $\sigma=0.001$                                      & \multicolumn{2}{c}{$\sigma=1.8$}                   \\
Random Jittering   & M=12                                               & M=5                                                 & \multicolumn{2}{c}{M=2}                            \\
Random Permutation & $\sigma=2$                                         & $\sigma=0.001$                                      & \multicolumn{2}{c}{$\sigma=2$}                     \\
    \bottomrule
    \end{tabular}
    }
    \label{tab:exp_datasets}
\end{table}


\subsection{Experimental Setup}
\label{exp:implementation}

\paragraph{Baseline Methods}
We consider two strands of self-supervised methods as baseline methods, namely \textit{contrastive-learning} based methods and \textit{masked autoencoding} based methods. Without loss of generality, for contrastive-based methods, we choose methods that utilize both positive and negative samples, SimCLR~\cite{chen2020simple}, SwAV~\cite{caron2020unsupervised}, and MoCo.V3~\cite{chen2021empirical}; we also consider the method that only requires positive sample pairs, BYOL~\cite{nips2020byol}. We compare the results of MoCo.V3 using both CNN and transformer architectures since it is compatible with both CNN and transformer architectures. Additionally, we pick TS-TCC~\cite{ijcai2021-324} and CoST~\cite{iclr2022CoST}, contrastive-based methods designed for time series pre-training as competitive baselines. We follow the data augmentations adopted in TS-TCC~\cite{ijcai2021-324} and choose random scaling, random jittering, and random permutation as data augmentations for contrastive learning-based approaches. \textcolor{black}{For the masked autoencoding-based methods, we choose GPT~\cite{chen2020generative} and \textcolor{black}{MaskedAE}~\cite{he2021masked}.} Besides the aforementioned methods, we also consider state-of-the-art supervised methods (as Sup. baselines), SimAttn~\cite{josephs2020semg} for EMG tasks and AttnSleep~\cite{eldele2021attention} for EEG tasks. We also provide supervised training results with random weight initialization of CNN and transformer backbones in Table \ref{tab:exp_network} for reference.

\paragraph{Evaluation Protocols}
\textcolor{black}{We evaluate Neuro-BERT using both the fine-tuning (FT) protocol~\cite{ICCV2019Scaling,he2021masked} and the linear probing protocol~\cite{ICCV2019Scaling}. The FT protocol, which holds significant practical value, is predominantly employed in real-world applications and focuses on harnessing the model’s strong non-linear features for superior performance on downstream tasks. Despite the emphasis on fine-tuning, we also report performance under the linear probing protocol, which is popular among contrastive learning researchers for demonstrating a model's linear discriminative capabilities. This protocol, however, does not fully exploit the non-linear potential of deep learning. 
For our experiments, each dataset is randomly split into training (80\%) and testing (20\%) subsets, with 20\% of the training data further allocated for validation. 
Table~\ref{tab:exp_datasets} presents the statistics and settings of all datasets. We conduct five trials for each task using different random seeds and report the mean and standard deviation for each metric. Specifically, for classification tasks, we assess top-1 accuracy (Acc), mean average precision (mAP), and macro-averaged F1-score (macro F1), which are particularly informative for evaluating performance in imbalanced datasets. For regression tasks, we report the mean absolute error (MAE) and mean squared error (MSE).
For the FT protocol, the network initialized with pre-trained weights is further fine-tuned on downstream tasks. The weights of the backbone, together with the prediction head for downstream tasks, are both updated. We fine-tune the full model for 40 epochs on labeled data for fully-supervised and semi-supervised scenarios. The training details are provided in Table~\ref{tab:app_ft}. Following linear protocols in contrastive learning~\cite{ICCV2019Scaling}, we evaluate the discrimination abilities of learned representations by training kNN classifiers ($k=20$) and linear SVMs~\cite{Boser1992svm} (\textit{one-vs-all}) on the frozen features.}


\begin{figure*}[t]
\centering
  \includegraphics[width=1.0\linewidth]{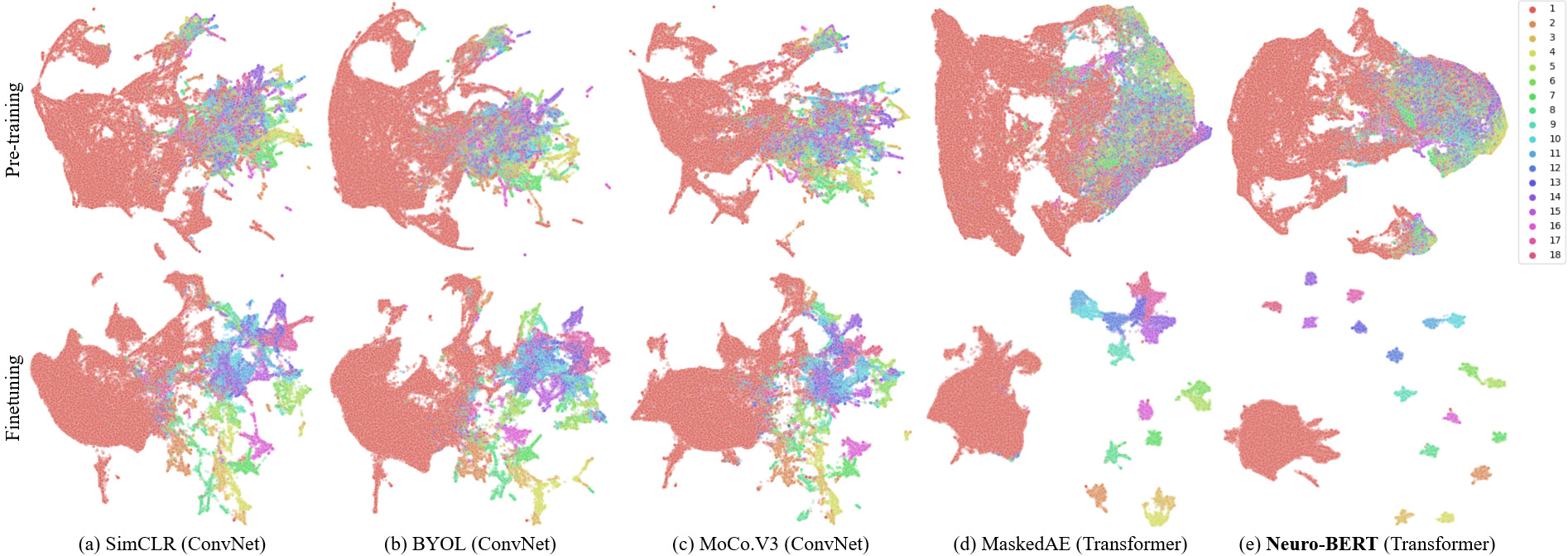}
  \vspace{-1.5em}
  \caption{Visualization of pre-trained and fine-tuned embeddings with UMAP of hand gesture classification task on the Ninapro dataset. The first row shows the visualization of embeddings after pre-training, while the second row shows the visualization of embeddings after fine-tuning. Contrastive learning-based methods demonstrate better discrimination ability after pre-training than \textcolor{black}{MaskedAE} and our proposed method. However, fine-tuning contrastive-based methods brings limited improvement, while masked autoencoding-based methods such as \textcolor{black}{MaskedAE} and our proposed methods gain huge improvement from fine-tuning.}
  \label{fig:exp_vis}
  \vspace{-1.0em}
\end{figure*}

\paragraph{Network Architecture}
\label{sec:network}
We adopt a simple yet effective plain convolution network architecture (ConvNet) as the backbone for all contrastive-based baseline methods, which is detailed in the left column of Tab.~\ref{tab:exp_network}. ConvNet contains a stem layer that contains a 1-D convolution layer with the kernel size of $K\times 1$, the stride of four, and the output channels of 32. The numbers of input channels $C$ and the kernel size $K$ are adjusted for each dataset, as shown in Tab.~\ref{tab:exp_datasets}. Each convolution block consists of a 1-D convolution layer (followed by a batch normalization layer and a ReLU activation) and a max pooling layer. For simplicity, we fix the convolution kernel size in each block and double the number of output channels concerning the previous block. Then, we adopt a global average pooling layer and a fully connected layer to map the output of the last block to fit the pre-training or the actual downstream tasks. Notice that all contrastive-based methods adopt an additional 2-layer MLP projector or predictor during pre-training as the original paper with the hidden dimension of 1024. For the transformer backbone, as illustrated in the right column of Tab.~\ref{tab:exp_network}, we perform patch embedding by a 1-D convolution layer with the kernel size of $P\times 1$ and the number of output channels of $128$, which is adjusted in each dataset as in Tab.~\ref{tab:exp_datasets}. The stride is also set to $P$ to ensure non-overlapping patch embedding~\cite{devlin2018bert,dosovitskiy2020image}. Following the stem layer are four blocks consisting of multi-head self-attention (MSA) with relative positional encoding (rel. pos.). We adopt the regular MSA structure with pre-normalization and residual connection as in ViT~\cite{dosovitskiy2020image}. The hidden dimension of the Feed Forward Network (FFN) is set to a fixed value of 512. We adopt the representation learned from the classification token and apply a fully connected layer for the pre-training or the actual downstream tasks in the case of the transformer. For impartial comparison, ConvNet and Transformer used in our experiments have a similar number of parameters in total (as shown in the last line of Tab.~\ref{tab:exp_network}) and yield similar performances.
\paragraph{Data augmentation.}
\label{data_augmentation}
We follow the data augmentations adopted in TS-TCC~\cite{ijcai2021-324} and choose random scaling, random jittering, and random permutation as data augmentations for contrastive learning-based approaches. Particularly, random scaling randomly changes the magnitude of the signals within a time window by multiplying a random scalar drawn from a Gaussian distribution with zero means and $\sigma_{s}$ as the standard deviation. Random jittering, on the other hand, could be considered as simulating additive noise. In particular, a random noise sampled from Gaussian distribution with zero means and $\sigma_{j}$ as the standard deviation is added to the original signal. Random permutation randomly perturbs the temporal ordering among consecutive time windows. More concretely, give a signal of a fixed time window, the signal is first split into $M$ segments with the same length. The ordering of the segments is randomly shuffled and concatenated to create a new signal. The hyperparameters $\sigma_s$, $\sigma_j$ and $M$ associated with data augmentations are listed in Tab.~\ref{tab:exp_datasets}. It is worth noticing that all mentioned data augmentations are utilized for contrastive learning-based approaches only, and no data augmentations are adopted for \textcolor{black}{MaskedAE} and our proposed Neuro-BERT.
\paragraph{Implementation Details}
\label{app:implement}
The experiments were performed using PyTorch~\cite{paszke2019pytorch} on NVIDIA V100 GPUs, with each experiment repeated five times with different random seeds. To ensure consistency with previous studies, we used the official codes provided by the authors of the baselines and followed their original configurations. Our evaluation metrics included mean square error (MSE) and mean absolute error (MAE) for regression tasks, top-1 accuracy (Acc), mean average precision (mAP), and macro-averaged F1-score (macro F1) and their average values for classification.
\begin{table}[t]
    \setlength{\tabcolsep}{0.9mm}
    \caption{Experimental setup for fine-tuning protocols based on both Transformer and CNN.}
    \centering
    \vspace{-0.5em}
\resizebox{0.99\columnwidth}{!}{
\begin{tabular}{c|c|c}
\toprule
Configuration          & Transformer                     & ConvNet                         \\ \hline
Fine-tuning epochs     & 40                              & 40                              \\
Optimizer              & AdamW                           & AdamW                           \\
Base learning rate     & 3e-4                            & 3e-4                            \\
Weight decay           & 0.01                            & 0.01                            \\
Optimizer momentum     & $\beta_1, \beta_2{=}0.9, 0.999$ & $\beta_1, \beta_2{=}0.9, 0.999$ \\
Batch size             & 128                             & 128                             \\
Learning rate schedule & cosine decay                    & cosine decay                    \\
Warmup epochs          & 5                               & -                               \\
Dropout ratio          & 0.2                             & 0.3                             \\
Gradient clipping      & 5.0                             & -                               \\
EMA                    & -                               & -                               \\
\bottomrule
\end{tabular}
    }
    \label{tab:app_ft}
\end{table}

\subsection{Comparison Results}
\label{exp:results}
\paragraph{Fine-tuning Results}
We first evaluate the performance of Neuro-BERT in comparison to current self-supervised methods with an end-to-end fine-tuning protocol (FT). As shown in Tab.~\ref{tab:exp_comp_ft}, Neuro-BERT significantly outperforms both the supervised and self-supervised pre-training methods for classification tasks on SleepEDF, Epilepsy, and Ninapro datasets. \textcolor{black}{Compared to contrastive-based methods, Neuro-BERT brings more fine-tuning performance gains over the random initialization than current state-of-the-art contrastive methods: +2.33\% Acc and +1.79\% macro F1 for Neuro-BERT with Transformer backbone v.s. +0.71\% Acc and +0.33\% macro F1 for TS-TCC with CNN backbone on SleepEDF, +2.31\% Acc and +5.23\% macro F1 for Neuro-BERT v.s. +0.18\% Acc and +1.77\% macro F1 for MoCo.V3 with Transformer backbone on Ninapro.} \textcolor{black}{Compared to the masked autoencoding-based pre-training approaches, Neuro-BERT noticeably outperforms the fine-tuning performances of both MaskedAE and GPT. In particular, Neuro-BERT outperforms GPT by 2.02\% in Acc and \textcolor{black}{3.56} \% in Macro F1 on Ninapro classification. It is worth noticing that GPT brings marginal performance gains over the random initialization compared to MaskedAE and Neuro-BERT. In natural language, the contextual information is rich and semantically dense, which facilitates the prediction of subsequent words based on preceding ones. In contrast, neurological signals are informatively sparse and stochastic, making it difficult to predict future signals based on past data. We hypothesize that this challenge hinders the effective extraction of generic features from neurological signals when using approaches like GPT, ultimately resulting in suboptimal performance for downstream tasks.} Meanwhile, Neuro-BERT also achieves the best performance for the finger joint angle regression task on the Ninapro dataset in Tab.~\ref{tab:exp_comp_ft}. Overall, our proposed Neuro-BERT achieves state-of-the-art performance in self-supervised neurological pre-training.
\paragraph{Linear Evaluation Results}
From Tab.~\ref{tab:exp_comp_linear}, we observe that contrastive-based methods outperform masked autoencoding-based methods due to their superior discrimination abilities induced by the natural clustering effect. However, contrastive-based methods perform way worse than masked autoencoding-based methods under FT. We argue that the discrimination ability comes with a cost of discarding essential information during pre-training, which subsequently restricts their performance enhancements during fine-tuning. On the contrary, masked autoencoding-based methods such as our proposed Neuro-BERT and MaskedAE capture low-level features and thus learn more general representations, which are more generic for various downstream tasks.

\paragraph{\textcolor{black}{Visualization Comparison of Learned Embeddings}} We provide visualization results of the representations learned after pre-training and fine-tuning for both contrastive learning-based approaches (SimCLR, BYOL, and MoCo.V3) and masked autoencoding-based methods (MaskedAE and Neuro-BERT), as shown in Fig.\ref{fig:exp_vis}. These visualizations are based on the hand gesture classification task on the Ninapro dataset, with the representations reduced to 2D using UMAP\cite{2018UMAP}. The experimental setup for this visualization follows the same procedures detailed in Sec.\ref{exp:implementation} and Tab.\ref{tab:app_ft}. As observed in Fig.~\ref{fig:exp_vis}, contrastive learning-based approaches demonstrate better linear separability after pre-training, an outcome of their inherent objectives to cluster similar samples and separate dissimilar ones, even without explicit class labels. In contrast, masked autoencoding-based methods capture a broader spectrum of low-level, generic features, though they do not necessarily induce a strong clustering effect initially. Notably, after fine-tuning, masked autoencoding-based methods exhibit significantly improved separability, outperforming the contrastive methods, which show only limited gains from fine-tuning. This pattern underscores the practical superiority of the fine-tuning protocol, particularly in its ability to enhance model performance significantly by leveraging the rich, general features acquired during pre-training.
\begin{figure}[tbp]
\centering
  \includegraphics[width=1.0\linewidth]{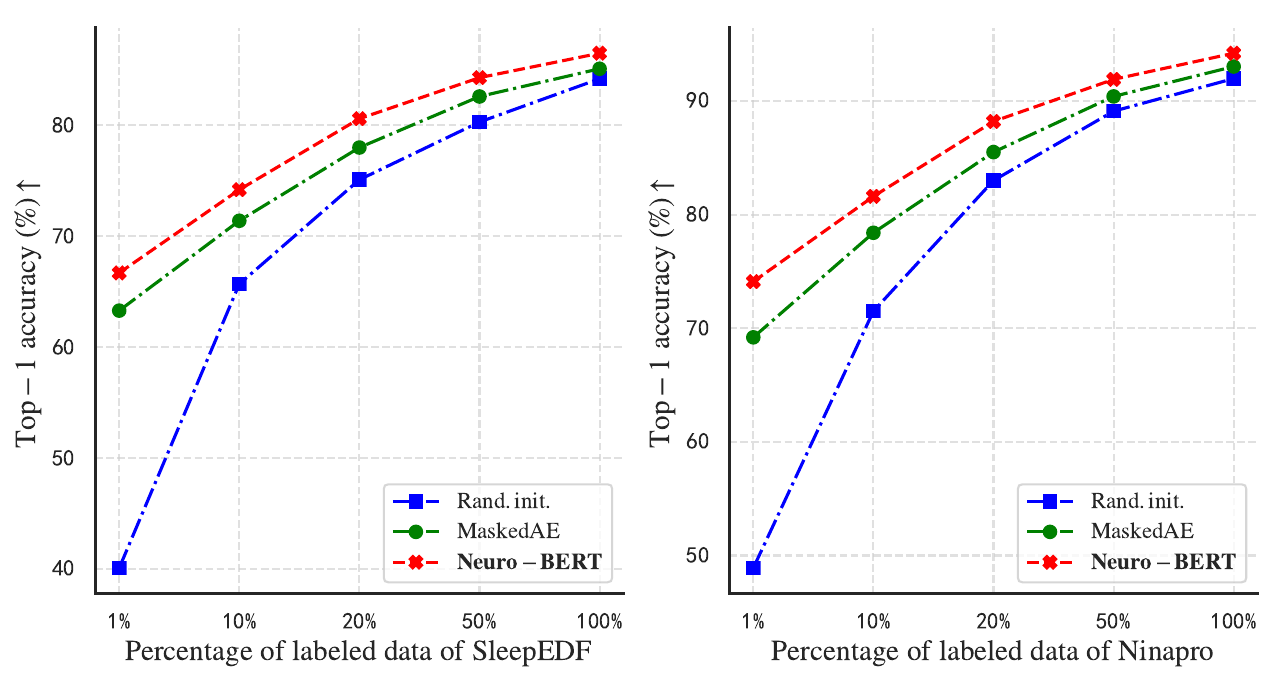}
  \vspace{-1.5em}
  \caption{Top-1 accuracy (\%) of fine-tuning protocol for the semi-supervised scenario on SleepEDF and Ninapro datasets.}
  \label{fig:exp_semi}
  \vspace{-1.0em}
\end{figure}

\paragraph{Semi-supervised Learning}
We then show that the proposed Neuro-BERT learns informative representations under semi-supervised learning settings. In particular, we randomly select 1\%, 10\%, 20\%, and 50\% labeled samples from the training dataset during the fine-tuning process. We report the top-1 accuracy of \textcolor{black}{MaskedAE}, Neuro-BERT, and random weight initialization on SleepEDF and Ninapro datasets. We use the same transformer structure as the backbone as described in Section \ref{sec:network} for all methods compared. As shown in Fig.~\ref{fig:exp_semi}, compared with the random initialization (blue curves) and \textcolor{black}{MaskedAE} (green curves), Neuro-BERT (red curves) steadily outperforms \textcolor{black}{MaskedAE} and random weight initialization on both SleepEDF and Ninapro datasets. We observe that the random initialization yields poor performance with limited labeled data (e.g., 1\% and 10\%). At the same time, our Neuro-BERT pre-training significantly improves the fine-tuning results in these cases (e.g., +30.31\% over the random initialization and +4.93\% over \textcolor{black}{MaskedAE} using 1\% labeled data on Ninapro).
\begin{figure}[t]
\centering
  \includegraphics[width=1.0\linewidth]{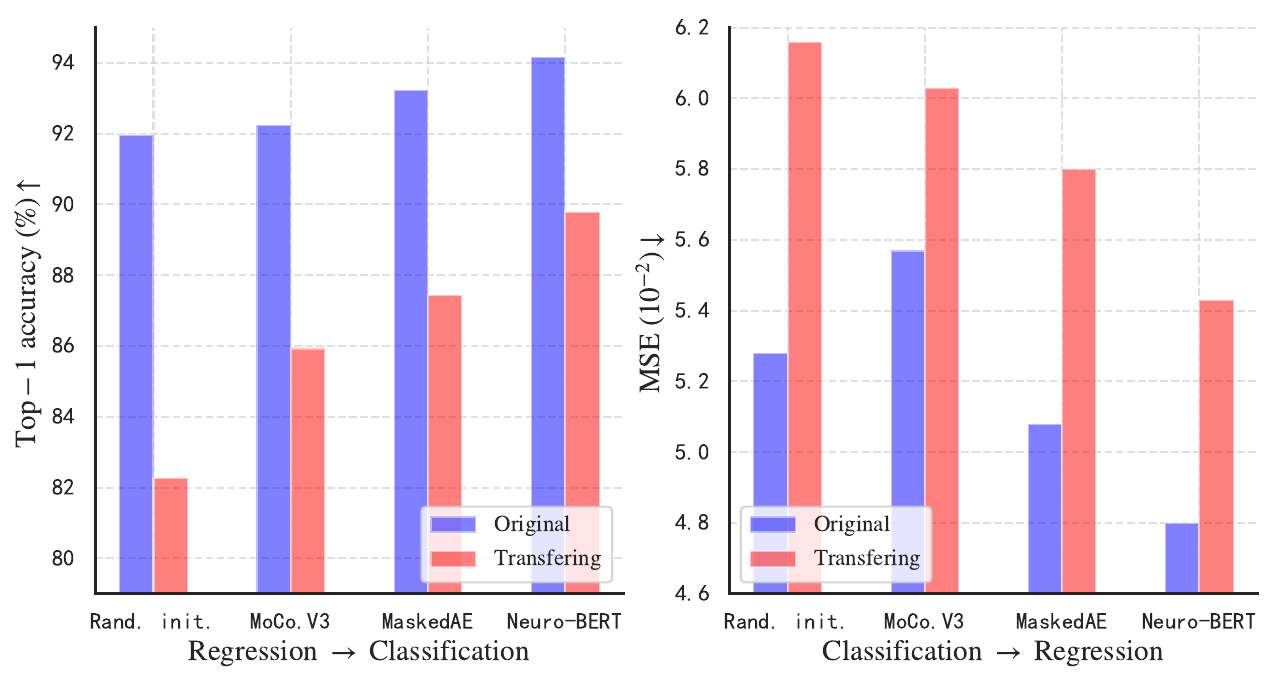}
  \vspace{-1.75em}
  \caption{Transfer learning performance comparison of pre-training between the classification and regression subsets of Ninapro datasets. Original indicates the performance of fine-tuning on the pre-trained weights with the targeted task labels directly without transferring.}
  \label{fig:exp_transfer}
  \vspace{-0.5em}
\end{figure}

\paragraph{Transfer Learning} We study the transferability of learned representations of Neuro-BERT via fine-tuning the pre-trained models in between the classification and regression tasks on the Ninapro dataset. We report the top-1 accuracy for classification and MSE for regression of random weight initialization, MoCo.V3, \textcolor{black}{MaskedAE}, and Neuro-BERT. The plotting on the left shows transfer learning from regression to classification, while the right shows vice versa. The blue bars show the original performance fine-tuned directly on the dataset with the targeted task label. As shown in Fig.~\ref{fig:exp_transfer}, Neuro-BERT yields the highest classification accuracy on the classification task and the lowest MSE on the regression task.
\begin{table}[t]
    \setlength{\tabcolsep}{1.3mm}
    \centering
    \caption{Ablation study of the decoder structures in terms of fine-tuning accuracy (\%) on the SleepEDF dataset.}
    \label{tab:ablation_advanced}
\resizebox{0.8\linewidth}{!}{
\begin{tabular}{c|c|c}
\hline
        & Module              & FT (Acc)$\uparrow$ \\ \hline
        & \gray{Linear}       & \gray{86.53$\pm$0.83}    \\
        & 2-layer MLP         & \textbf{86.62}$\pm$\textbf{0.74}               \\
Decoder & 4-layer MLP         & 86.19$\pm$0.69               \\
        & 2-layer Transformer & 86.48$\pm$0.87               \\
        & 4-layer Transformer & 86.04$\pm$0.78               \\ \hline
\end{tabular}
    }
\end{table}

\subsection{Ablation Studies}
\label{exp:ablation}
\begin{figure}[t]
\centering
  \includegraphics[width=0.75\linewidth]{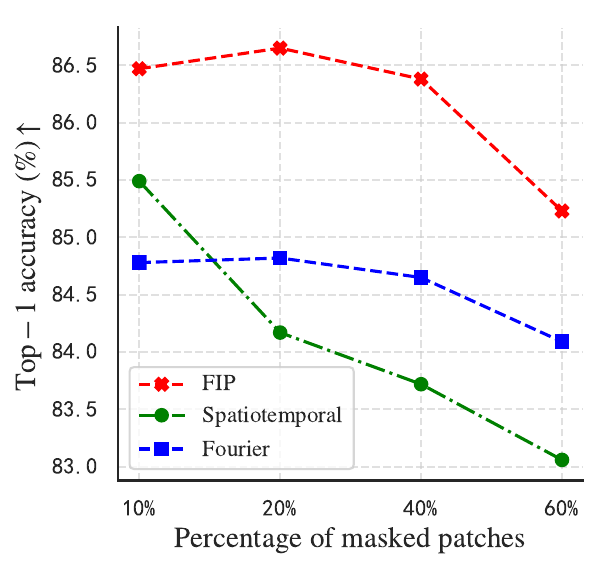}
  \caption{Ablation study of the masking ratios in both the spatiotemporal and the Fourier domains in terms of fine-tuning accuracy (\%) on the SleepEDF dataset.}
  \label{fig:exp_ablation}
\end{figure}

\textcolor{black}{This section ablates three key designs: masking ratio, prediction target, and decoder design. Ablation studies are performed on the SLEEP-EDF dataset with the same experimental setup as described in Section \ref{exp:implementation}. We report 40-epoch fine-tuning accuracy (\%) on SLEEP-EDF. Additionally, we examined the applicability of the proposed FIP across other masked autoregressive modeling approaches, including GPT and MaskedAE.}
\paragraph{Ablation on Masking Ratio and Prediction Target}The masking ratios and prediction targets play important roles in masked autoencoding-based self-pretraining approaches. In this section, we study the effects of the masking ratios with respect to different prediction targets. For the prediction targets, we consider both the spatiotemporal domain and Fourier domain targets, namely, directly predicting the missing signal in the spatiotemporal domain (spatiotemporal), predicting the Fourier spectrum (Fourier), and the proposed FIP. As observed in Fig. \ref{fig:exp_ablation}, we discover that the best performance is given with a masking ratio of 10\% on the spatiotemporal domain target. A drastic performance drop is also observed from the 10\% masking ratio to the 60\% making ratio. This indicates that neurological signals are informatively sparse in the spatiotemporal domain, and thus, performing masked autoencoding prediction in the spatiotemporal domain would be a very improper and redundant task as modeling the stochasticity and nonstationarity of neurological signals. On the contrary, adopting Fourier-domain prediction targets shows better robustness and accuracy performance in general. For FIP, we observe that masking ratios of 10\%, 20\% to 40\% can produce similar fine-tuning accuracy. For directly predicting the Fourier spectrum as the target, we observe that the performance is relatively stable under various masking ratios and worse performance than FIP. Given the observed results, our proposed FIP of Neuro-BERT outperforms the performance of masked autoencoding prediction in the spatiotemporal domain. More importantly, the performance of Neuro-BERT is robust against different masking ratios.
\begin{table}[t]
    \setlength{\tabcolsep}{1.3mm}
    \centering
    \caption{Ablation study of the proposed FIP integrated into other autoregressive pre-training approaches. Fine-tuning accuracy (\%) on the SleepEDF dataset is reported.}
    \label{tab:ablation_FIP}
\resizebox{0.6\linewidth}{!}{
\begin{tabular}{c|c}
\hline
Method       & FT (Acc)$\uparrow$  \\ \hline
\gray{Neuro-BERT}   & \gray{86.53$\pm$0.83} \\
GPT~\cite{chen2020generative}          & 84.56$\pm$0.77     \\
GPT+FIP      & 85.03$\pm$0.65      \\
MaskedAE~\cite{he2021masked}     & 85.08$\pm$0.89      \\
MaskedAE+FIP & 85.94$\pm$0.82      \\ \hline
\end{tabular}
    }
\end{table}
\paragraph{Ablation on Decoder Design} Next, we conduct an ablation study to assess the impact of various decoder designs on our model's performance. By default, Neuro-BERT utilizes a linear layer decoder. To explore alternative architectures, we also evaluate Multi-Layer Perceptrons (MLPs) and Transformers with varying numbers of layers. For MLPs, the hidden dimension is set to 1024. We adopt the same Transformer block design as described in Tab.~\ref{tab:exp_network}. As observed in Tab.~\ref{tab:ablation_advanced}, {}we found that adopting a 2-layer MLP offers marginal improvements compared to a linear decoder. Overly complex decoder structures such as 4-layer MLP and Transformer decoders tend to degrade performance.
\paragraph{Effectiveness Verification of FIP} we further investigate the effectiveness of our \textbf{Fourier Inversion Prediction (FIP)} across other masked autoregressive modeling approaches, including GPT and MaskedAE. The corresponding fine-tuning classification results on the SLEEP-EDF dataset are shown in Tab.~\ref{tab:ablation_FIP}. Notably, the integration of FIP into MaskedAE results in a substantial performance boost, while the improvements with GPT are relatively modest. This observation reinforces our prior assertion that the strategy of predicting future signals from past signals is typically unsuitable for neurological data due to their inherent stochasticity and nonstationarity. Nevertheless, FIP offers a significant enhancement over the conventional next-token style time-domain reconstruction method utilized by GPT, effectively alleviating some of GPT’s shortcomings and enhancing its performance. It is also important to note that, unlike MaskedAE, which only processes visible tokens, our proposed Neuro-BERT architecture inputs both visible and masked tokens into the encoder. This further demonstrates the versatility of FIP as a universally applicable enhancement for masked autoencoding-based neurological pretraining.
\label{ablation:decoder}
\section{Discussion and Future Work}
Learning neurological representations that are generic for downstream tasks in a self-supervised manner is a promising research direction for the next generation of BCI with the rise of the concept of the metaverse. However, existing approaches tend to directly adopt well-established pre-training methods (contrastive learning) from the CV domain, neglecting the uniqueness of neurological signals. We take a step forward to design a masked autoencoding-based pre-training framework. Our intuition is based on the biomedical foundation that frequency and phase distribution of neurophysiology signals reveal the underlying neurological activities. Thus, recovering the masked signal via IDFT of the predicted frequency and phase brings more explainability. This design also enables our proposed Neuro-BERT to yield a decent performance with only one linear layer as the prediction head as opposed to a complex structure transformer utilized by MAE. We hope that our exploration of prediction targets in terms of self-supervised masked autoencoding neurological representation provides insights to the community.

Next, we list three potential limitations of Neuro-BERT: (i) Although Neuro-BERT does not require data annotation, Neuro-BERT still relies on abundant unlabeled data for representation learning. Neuro-BERT yields limited performance gain where only a small dataset is available. This is also a drawback of most self-supervised pre-training methods; (ii) Neuro-BERT cannot deal with the case where the input neurological signals are noisy. Since the noise will be present in the Fourier domain more or less, Nuero-BERT will also be forced to model noise if the frequency magnitude of the noise is high; (iii) Compared to contrastive learning approaches, the linear separability of the learned representations is not as good. Although linear separability is not crucial for real applications, it is also desired to improve the linear separability of the representations learned by Nuero-BERT.

In future work, we plan to tackle the aforementioned limitations: (i) Design an adaptive weighting mechanism for the frequency loss function in the Fourier domain for the masked autoencoding task. It is desired that the frequencies associated with noise can be weighted down by the adaptive frequency loss function; (ii) Combine contrastive learning with masked autoencoding. The learned representations incorporate both the advantages of contrastive learning and masked autoencoding; (iii) Extend our proposed framework to a multi-modal framework containing images, text, and neurological signals to improve user interaction experience with computers further.

\section*{References}
\vspace{-3mm}
{
\bibliographystyle{ieeetr}
\bibliography{ref.bib}
}
\end{document}